\documentstyle[12pt]{article}

\newcommand\M{{\cal M}}
\newcommand\OO{{\cal O}}
\newcommand\op{\omega_{\rm P}}
\newcommand\kmax{k_{\rm max}}
\newcommand\half{{\textstyle{1\over2}}}

\newcommand\lapprox{\mathrel{\mathop
  {\hbox{\lower0.5ex\hbox{$\sim$}\kern-0.8em\lower-0.7ex\hbox{$<$}}}}}
\newcommand\gapprox{\mathrel{\mathop
  {\hbox{\lower0.5ex\hbox{$\sim$}\kern-0.8em\lower-0.7ex\hbox{$>$}}}}}

\begin{document}

\title{Standard and Non-Standard Plasma\\
Neutrino Emission Revisited}

\author{Martin Haft and Georg Raffelt\\
Max-Planck-Institut f\"ur Physik \\
F\"ohringer Ring 6, 80505 M\"unchen, Germany\\
\\
Achim Weiss\\
Max-Planck-Institut f\"ur Astrophysik\\
85748 Garching, Germany}

\maketitle

\begin{abstract}
On the basis of Braaten and Segel's representation of the electromagnetic
dispersion relations in a QED plasma we check the numerical accuracy of
several published analytic approximations to the plasma neutrino emission
rates. As we find none of them satisfactory we derive a new analytic
approximation which is accurate to within 4\%\ where the plasma process
dominates. The correct emission rates in the parameter regime relevant for
the red giant branch in globular clusters are larger by about $10-20$\% than
those of previous stellar evolution calculations.  Therefore, the core mass
of red giants at the He flash is larger by about $0.005\M_\odot$ or 1\% than
previously thought. Our bounds on neutrino magnetic dipole moments remain
virtually unchanged.
\end{abstract}

\vfil

\begin{center}
To appear in the {\it Astrophysical Journal}
\end{center}

\eject

\section{Introduction}
In stars with densities below nuclear, neutrinos are emitted by the plasma
process $\gamma\to\overline\nu\nu$, the photo process
$\gamma e^-\to e^-\overline\nu\nu$, the pair process
$e^-e^+\to\overline\nu\nu$, and by bremsstrahlung
$e^-(Ze)\to (Ze)e^-\overline\nu\nu$. In Fig.~1 we show the regions of density
and temperature where each of these processes dominates over the others. The
plasma process is particularly interesting because it does not occur in vacuum
and yet dominates the stellar emission rates in a large range of temperatures
and densities. Notably white dwarfs and the degenerate cores of low-mass red
giants fall in this parameter region. Evolutionary calculations for these
stars can be tested with great statistical significance. For example, globular
cluster observations allow one to determine the core mass at the helium flash
to within about $0.012\,\M_\odot$ or 2.5\% at a $1\sigma$ statistical
confidence level (Raffelt 1990, Raffelt and Weiss 1992). Therefore, it is
important to identify possible systematic effects that would enter at this or
at a larger level. If the standard neutrino loss rates are multiplied with a
factor $F_\nu$ where $F_\nu=1$ refers to the standard case, the core mass at
the helium flash varies approximately as
$\delta\M_{\rm c}=0.020\,\M_\odot\,\delta F_\nu$. This strong sensitivity to
the neutrino luminosity has been used to constrain possible non-standard
contributions such as the plasmon decay by virtue of neutrino dipole moments
(Raffelt 1990, Raffelt and Weiss 1992).

In practical stellar evolution calculations the neutrino loss rates are
implemented in the form of simple analytic approximations. Widely used
versions are those of Beaudet, Petrosian, and Salpeter (1967), Munakata,
Kohyama, and Itoh (1985), Schinder et~al.\ (1987), and Itoh et~al.\ (1989).
With regard to the plasma process, the results of Beaudet et~al., Munakata
et~al., and Schinder et~al.\ agree with each other to better than 1\%\ if the
same neutrino coupling strength to electrons is used, while Itoh et~al.\ made
an attempt to improve the accuracy of the rates at low temperatures.

It turns out that all of these rates are relatively poor approximations for
$T\lapprox10^8\,K$ which is relevant for low-mass stars. They were optimized
for higher temperatures and correspondingly higher densities in the diagonal
band of Fig.~1 where the plasma process dominates. At higher densities,
however, the approximate photon dispersion relation that had been used in all
of these works is a rather poor approximation (Braaten 1991) with the result
that the above plasma emission rates are bad approximations everywhere. In
response, Itoh et al.\ (1992) have derived a new analytic formula which is
claimed to fit the exact results to better than 5\%\ in the region where the
plasma process dominates. Independently, Blinnikov and Dunina-Barkovskaya
(1992) have published rates which were optimized for
low-mass stars.

Since then Braaten and Segel (1993) have devised an approach to the photon
dispersion relation which, for given $T$ and $\rho$, reduces the calculation
of the plasma emission rate essentially to the numerical evaluation of a few
one-dimensional integrals. Their method for the first time offers a simple and
practical approach to check the accuracy of all of the above emission rates as
well as the approximations that were used to study non-standard neutrino
emission (Raffelt 1990, Raffelt and Weiss 1992, Blinnikov and
Dunina-Barkovskaya 1992, Castellani and Degl'Innocenti 1992). On the basis of
this method we found none of the above approximations satisfactory and thus
have derived a new analytic approximation to the standard rate for the plasma
process which is accurate to better than 4\%.  For the non-standard emission
rates we found that a simple scaling of the standard rates as in Raffelt and
Weiss (1992) introduces only a small error (less than 5\%) for low-mass stars
so that bounds on neutrino dipole moments are mostly affected by the accuracy
of the standard emission rates.

We begin in Sect.~2 by adapting the results of Braaten and Segel (1993) to our
calculation of standard and non-standard emission rates by the plasma process.
Besides the standard neutrino interactions we include the possibility of the
coupling of right-handed neutrinos to electrons (Fukugita and Yanagida 1990),
neutrino dipole moments, and neutrino electric ``millicharges''. In Sect.~3 we
perform a numerical test of all of the above analytic approximations, and we
derive and test a new version. We also check the accuracy of a simple scaling
of the standard rates to obtain the non-standard ones. In Sect.~4 we discuss
the accuracy of previous calculations of the core-mass at the helium flash as
well as the accuracy of previous bounds on neutrino properties, and we derive
bounds on the Fukugita and Yanagida model as well as on neutrino millicharges.

\section{Exact Emission Rates}

\subsection{Electromagnetic Excitations in a Medium}

The behavior of electromagnetic excitations in a linear medium is best
understood on the basis of the wave equation for the vector potential $A$ in
Feynman gauge, $[K^2-\Pi]\,A=j_{\rm ext}$, where $j_{\rm ext}$ is an external
electromagnetic current and $\Pi$ is the polarization tensor.  Moreover,
$K^2=\omega^2-k^2$ where $\omega$ is the frequency while $k=\vert{\bf k}\vert$
the modulus of the momentum coordinate.  In an isotropic medium which also
remains invariant under a parity transformation, the polarization tensor can
be expressed as (Nieves and Pal 1989a,b)
$\Pi=\pi_T Q_T+\pi_L Q_L$ where $Q_T$ is a $K$-dependent projector on the
subspace transverse to $\bf k$ while $Q_L$ projects on the longitudinal
subspace, and in addition $Q_{T,L}K=0$ so that $\Pi\,K=0$ as required by gauge
invariance. Explicit expressions for $Q_{T,L}$ in terms of $K$ were given, for
example, by Weldon (1982). He also showed that in the rest-frame of the medium
$\pi_L=-\Pi_{00}\,K^2/k^2$ and
$\pi_T=(\Pi^\mu_\mu-\pi_L)/2$.

If $\Pi$ is expressed in this way, it is clear that the homogeneous wave
equation $(K^2-\Pi)\,A=0$ has non-trivial solutions only for
\begin{equation}
\omega^2-k^2-\pi_s(\omega,k)=0,\qquad\hbox{$s=T$ or $L$,}
\label{Xc}
\end{equation}
a relationship which implicitly defines the dispersion relations for
transverse and longitudinal propagating modes (``photons'' and ``plasmons'' or
``transverse and longitudinal plasmons'').

In order to calculate the decay rate into neutrinos of such modes we need to
normalize properly the amplitude of the quantized excitations. To this end we
consider a mode $k$, polarization $s=T$ or $L$, with the corresponding
frequency $\omega_{s,k}$ so that the dispersion relation Eq.~(\ref{Xc}) is
obeyed. Then we expand
$\pi_s(\omega,k)=\pi_s(\omega_{s,k},k)+\partial_{\omega^2}\pi_s
(\omega_{s,k},k) (\omega^2-\omega_{s,k}^2)$ so that the equation of motion for
the amplitude $A_{s,k}$ is that of a harmonic oscillator. Equivalently, we
write the photon propagator near its poles in the form $Z_s/(\omega^2-
\omega^2_{s,k})$. Either way, one finds that the squared matrix element of a
process with an external electromagnetic excitation of momentum $k$ and
polarization $s$ carries a factor
\begin{equation}
Z_s=[1-\partial_{\omega^2}\pi_s(\omega_{s,k},k)]^{-1}\,.
\label{Xd}
\end{equation}

In a fully ionized plasma the polarization tensor is simply given in terms of
the forward scattering amplitudes of photons on the electrons and positrons.
To lowest order in $\alpha=e^2/4\pi$ it is found to be \begin{equation}
\Pi_{\mu\nu}=-8\pi\alpha\int\frac{d^3{\bf p}}{(2\pi)^3}
\,\frac{f(E_p)}{E_p}\,
\frac{P{\cdot}K\,(K_\mu P_\nu+K_\nu P_\mu)-K^2 P_\mu P_\nu-
(P{\cdot}K)^2 g_{\mu\nu}}{(P{\cdot}K)^2-K^4/4}
\label{Xe}
\end{equation}
where $P=(E_p,{\bf p})$ is an $e^-$ or $e^+$ four-momentum. The sum of the
phase-space occupation numbers for these particles is
\begin{equation}
f(E)=\frac{1}{e^{(E-\mu)/T}+1}+\frac{1}{e^{(E+\mu)/T}+1}
\end{equation}
with $\mu$ the electron chemical potential and $T$ the temperature. On the
basis of Eq.~(\ref{Xe}) it is, in principle, straightforward to derive
$\pi_s$ and $Z_s$ for a given $\mu$ and $T$.

\subsection{Analytic Representation for $\Pi$ and $Z$}
Braaten and Segel (1993) have shown that ignoring the $K^4/4$ term in the
denominator of Eq.~(\ref{Xe}) introduces an error which appears only on the
$\alpha^2$ level so that to $\OO(\alpha)$ it can be ignored. In fact,
ignoring this term provides a {\it better\/} approximation to the
$\OO(\alpha)$ result than keeping it because $\pi_T$ then remains real for all
conditions as it must because the imaginary part from the $\gamma_T\to e^+e^-$
process which otherwise appears at sufficiently large plasma frequencies is
unphysical (Braaten 1991).

Once the $K^4/4$ term has been dropped, the angular part of the integral in
Eq.~(\ref{Xe}) can be done analytically, leaving one with a one-dimensional
integral over electron momenta which can be done analytically in the
classical, degenerate, and relativistic limit. In these cases Braaten and
Segel found\footnote{Braaten and Segel used Coulomb rather than Feynman gauge
so that we had to translate their longitudinal expression accordingly. We
preferred to follow Weldon (1982) in the choice of gauge because the
dispersion relations and plasmon decay rates then have the same form for
transverse and longitudinal excitations.}
\begin{eqnarray}
\pi_T(\omega,k)&=&\op^2\Bigl[1+{\textstyle{1\over 2}} G(v_*^2
k^2/\omega^2)\Bigr]\,, \nonumber\\
\pi_L(\omega,k)&=&\op^2\Bigl[1-G(v_*^2 k^2/\omega^2)\Bigr]
+v_*^2k^2-k^2 \,,
\label{Xf}
\end{eqnarray}
where $v_*\equiv\omega_1/\op$ has the interpretation of a typical velocity of
the electrons in the medium. The plasma frequency $\op$ and the frequency
$\omega_1$ are defined by
\begin{eqnarray}
\op^2&\equiv&\frac{4\alpha}{\pi}\int_0^\infty dp\,
(v-{\textstyle{1\over3}}v^3)\,p\, f(E_p)\,,\nonumber\\
\omega_1^2&\equiv&\frac{4\alpha}{\pi}\int_0^\infty dp\,
({\textstyle{5\over3}} v^3-v^5)\,p\,f(E_p)\,,
\label{Xff}
\end{eqnarray}
where $v=p/E_p$ is the electron or positron velocity. In the classical limit
one has $v_*=(5T/m_e)^{1/2}$, in the degenerate limit $v_*=v_{\rm F}$
(velocity at the Fermi surface), and in the relativistic limit $v_*=1$. The
function $G$ is defined by
\begin{equation}
G(x)\equiv\frac{3}{x}\left[1-\frac{2x}{3}-\frac{1-x}{2\sqrt{x}}
\log\frac{1+\sqrt{x}}{1-\sqrt{x}}\right]
=6\sum_{n=1}^\infty\frac{x^n}{(2n+1)(2n+3)}\,.
\label{Xfff}
\end{equation}
We have plotted $G(x)$ in Fig.~2; note that $G(0)=0$, $G(1)=1$, and
$G'(1)=\infty$.

Braaten and Segel then claim (and we have checked) that these results apply
approximately for all conditions. The deviations between $\pi_s(\omega,k)$
given by Eq.~(\ref{Xf}) and by the proper integral over the $e^+e^-$ phase
space are always so small that Eq.~(\ref{Xf}) can be considered exact to
$\OO(\alpha)$. As a higher precision would require an $\OO(\alpha^2)$
QED
calculation, nothing is gained by evaluating the full integrals. Therefore,
the above results are as exact as an $\OO(\alpha)$ result can be.

In order to calculate the plasmon decay rates we also need the amplitude
normalization factors. Braaten and Segel (1993) found the analytic
representations
\begin{eqnarray}
Z_{T,k}&=&\frac{2\omega_{T,k}^2(\omega_{T,k}^2-v_*^2k^2)}{\omega_{T,k}^2
(3\op^2-2\pi_{T,k})+(\omega_{T,k}^2+k^2)(\omega_{T,k}^2-v_*^2k^2)}\,,
\nonumber\\
Z_{L,k}&=&\frac{2\omega_{L,k}^2 (\omega_{L,k}^2-v_*^2k^2)}{[3\op^2
-(\omega_{L,k}^2-v_*^2k^2)] \pi_{L,k}}\,.
\label{Xffff}
\end{eqnarray}
In these expressions $\pi_{s,k}\equiv\pi_s(\omega_{s,k},k)$ is the ``on
shell'' value of $\pi_s$ for excitations with momentum $k$, i.e.\
$\omega_{s,k}^2-k^2=\pi_{s,k}$.

The decay $\gamma_s\to\overline\nu\nu$ is only possible if $\gamma_s$ has a
time-like four-momentum, $\omega_{s,k}>k$. For longitudinal excitations this
condition is only fulfilled for $k<\kmax$ where
\begin{equation}
\kmax=\op\,\left[\frac{3}{v_*^2}\left(\frac{1}{2v_*}
\log\frac{1+v_*}{1-v_*}-1\right)\right]^{1/2}
\label{Xg}
\end{equation}
to the same level of approximation.

In the Braaten and Segel representation the on-shell values $\pi_{s,k}$
and thus the dispersion relations depend only on the medium parameters $\op$
and $v_*$. However, media are more readily characterized by their density and
temperature. Therefore, in Fig.~3 we show contours of constant $v_*$ and
$\gamma\equiv\op/T$ in the region of $\rho$ and $T$ relevant for the plasma
process in stars. In the shaded area the plasma process contributes more than
10\% to the total neutrino luminosity. Evidently, it is important only for
$0.3\lapprox\gamma\lapprox30$. Moreover, the $v_*$ contours are almost
perfect vertical straight lines in this regime, i.e.\ the medium is
degenerate.

The polarization functions $\pi_{s,k}/\op^2$ and the amplitude factors
$Z_{s,k}$ are functions only of $v_*$ and of $k/\op$. In Fig.~4 we show
contours for these functions in the $v_*$-$k$-plane for transverse
excitations. We have $1<\pi_{T,k}/\op^2<3/2$ and $Z_{T,k}<1$. The deviation of
$Z_{T,k}$ from unity is always small. In Fig.~5 we show similar contours for
the longitudinal case. We have $\pi_{L,k}/\op^2<1$; the occurrence of the
plasma process in addition requires $0<\pi_{L,k}$. The contour $\pi_{L,k}=0$
is identical with the function $\kmax(v_*)$ given in Eq.~(\ref{Xg}). The
amplitude function $Z_{L,k}$ diverges when $k\to\kmax$.  Therefore, we show in
Fig.~5 (lower panel) contours for
$Z_{L,k}^*\equiv Z_{L,k}\pi_{L,k}/\omega_{L,k}^2$
instead.\footnote{Our $Z_{L,k}^*$ is what Braaten and Segel (1993) call
$Z_{L}(k)$.}

\subsection{Plasmon Decay Rates}

In order to calculate the neutrino emission rates for the standard and several
non-standard interaction models it remains to calculate the decay rates for
the process $\gamma_s\to\overline\nu\nu$. In the standard model one finds
(Adams, Ruderman and Woo 1963, Zaidi 1965, Dicus 1972)
\begin{equation}
\Gamma_{s,k}=C_V^2\,\frac{G_{\rm F}^2}{48\pi^2\alpha}\,
\frac{Z_{s,k}\,\pi_{s,k}^3}{\omega_{s,k}}\,.
\label{Ya}
\end{equation}
The effective vector coupling constant is
\begin{equation}
C_V^2\equiv\sum_{i=1}^3 C^2_{V,i}=(\half+2\sin^2\Theta_{\rm W})^2 +2(\half-
2\sin^2\Theta_{\rm W})^2\,,
\label{Yb}
\end{equation}
where $C_{V,i}$ is the effective neutral-current vector coupling constant of
neutrino flavor $i$ to the electrons. With a weak mixing angle of
$\sin^2\Theta=0.2325\pm0.0008$ this is
$C_V^2=(0.9312\pm0.0031)+2\,(0.0012\pm0.0001)=0.9325\pm0.0033$ where the first
term is from $\gamma_s\to\overline\nu_e\nu_e$ while the second term is from
$\overline\nu_\mu\nu_\mu$ and $\overline\nu_\tau\nu_\tau$. Thus, even if these
latter flavors were too heavy to be produced by plasmon decay the value of
$C_V^2$ would change by less than its uncertainty. The contribution of the
axial neutral current is always negligible.

As a first non-standard coupling we consider a model by Fukugita and Yanagida
(1990) which was devised to obtain a large neutrino magnetic dipole moment and
a strong effective coupling of right-handed $\nu_e$'s to electrons. The
relevant part of the Lagrangian is
\begin{equation}
{\cal L}_{\nu_R e_L}=g\phi \overline\psi_{\nu_R}\psi_{e_L}
+{\rm h.c.}
\label{Yc}
\end{equation}
where $\phi$ is a scalar field of mass $M$ and $g$ is a dimensionless coupling
constant. For low energies this interaction produces an effective
neutral-current coupling of r.h.\ neutrinos to electrons. The corresponding
effective vector coupling constant is
\begin{equation}
C_{V,R}^2=\frac{g^4}{32 M^4 G_{\rm F}^2}\,.
\label{Yd}
\end{equation}
In this model the rate for $\gamma_s\to\overline\nu_{e,R}\nu_{e,R}$ is found
by inserting $C_{V,R}^2$ instead of $C_V^2$ into Eq.~(\ref{Ya}).

Next, we consider direct couplings of neutrinos with the electromagnetic
field. The least exotic possibility is that of neutrino dipole moments with an
effective Lagrangian
\begin{equation}
{\cal L}_\mu=\frac{1}{2}\sum_{i,j=1}^3
\Bigl(\mu_{ij}\overline\psi_i\sigma_{\mu\nu}\psi_j+
\epsilon_{ij}\overline\psi_i\sigma_{\mu\nu}\gamma_5\psi_j\Bigr)
\,F^{\mu\nu}\,,
\label{Ye}
\end{equation}
where $\mu_{ij}$ and $\epsilon_{ij}$ are matrices of magnetic and electric
dipole and transition moments, $F$ is the electromagnetic field tensor, and
the sum is over neutrino flavors. We define an effective dipole moment by
\begin{equation}
\mu^2_\nu\equiv\sum_{i,j=1}^3\Bigl(\vert\mu_{ij}\vert^2
+\vert\epsilon_{ij}\vert^2\Bigr)
\label{Yf}
\end{equation}
with the restriction that the sum should only run over those flavors which are
light enough to be produced by the plasma process: $m_i+m_j\lapprox\op$. Then
we find for the plasmon decay rate
\begin{equation}
\Gamma_{s,k}=\frac{\mu_\nu^2}{24\pi}\,
\frac{Z_{s,k}\,\pi_{s,k}^2}{\omega_{s,k}}\,.
\label{Yg}
\end{equation}
This result is in agreement with Sutherland et al. (1976) except for their
$Z_{L,k}$.

Finally, we assume that neutrino flavor $i$ carries a ``millicharge'' $q_i e$.
In this case we find
\begin{equation}
\Gamma_{s,k}=\frac{q_\nu^2\alpha}{3}\,
\frac{Z_{s,k}\,\pi_{s,k}}{\omega_{s,k}}\,,
\label{Yh}
\end{equation}
where $\alpha=e^2/4\pi$ is the (electron) fine structure constant and
$q_\nu^2\equiv\sum_{i=1}^3q_i^2$.

\subsection{Neutrino Emission Rates}

If transverse and longitudinal electromagnetic excitations of momentum $k$ can
decay according to $\gamma_s\to\overline\nu\nu$ with a rate $\Gamma_{s,k}$
then the energy-loss rate per unit volume of a medium at temperature $T$ is
$Q=Q_T+Q_L$ where
\begin{eqnarray}
Q_T&=&2\int_0^\infty \frac{dk\,k^2}{2\pi^2}\,
  \frac{\Gamma_{T,k}\, \omega_{T,k}}{e^{\omega_{T,k}/T}-1}\,,
\nonumber\\
Q_L&=&\int_0^{\kmax}\frac{dk\,k^2}{2\pi^2}\,
  \frac{\Gamma_{L,k}\,\omega_{L,k}}{e^{\omega_{L,k}/T}-1}\,.
\end{eqnarray}
Inserting the results of the previous section into this equation we find
for the various interaction models
\begin{eqnarray}
Q_{V}&=&C_V^2\,\frac{G_{\rm F}^2}{96\pi^4\alpha}\,T^3\,
\op^6\,Q_3\,,\nonumber\\
Q_\mu&=&\frac{\mu_\nu^2}{48\pi^3}\,T^3\,\op^4\,Q_2\,,\nonumber\\
Q_q&=&\frac{q_\nu^2\alpha}{6\pi^2}\,T^3\,\op^2\,Q_1\,.
\label{YYa}
\end{eqnarray}
The dimensionless emission rates are $Q_n\equiv(Q_n^T+Q_n^L)$ where
\begin{eqnarray}
Q^T_n&\equiv&2\int_0^\infty \frac{dk\,k^2}{T^3}\,Z_{T,k}\,
      \left(\frac{\pi_{T,k}}{\op^2}\right)^n
      \frac{1}{e^{\omega_{T,k}/T}-1} \,,\nonumber\\
Q^L_n&\equiv&\int_0^{\kmax}\frac{dk\,k^2}{T^3}\,Z_{L,k}\,
      \left(\frac{\pi_{L,k}}{\op^2}\right)^n
      \frac{1}{e^{\omega_{L,k}/T}-1} \,.
\label{YYb}
\end{eqnarray}
In the Braaten and Segel approximation the $Q^s_n$ are only functions of $v_*$
and $\gamma=\op/T$.

In Fig.~6 we show contours in the $v_*$-$\gamma$-plane of $Q^L_3/Q^T_3$, i.e.\
the ratio between longitudinal and transverse emissivity for the
standard-model interactions. Corresponding contours in the $\rho$-$T$-plane
are shown as dashed lines in Fig.~1. Evidently, the longitudinal process is of
importance only in a relatively narrow region near $\gamma=10$.

In a practical calculation of anomalous neutrino losses one will scale the
standard luminosity appropriately. The relevant ratios are
\begin{equation}
\frac{Q_\mu}{Q_V}=\frac{\mu_\nu^2\,2\pi\alpha}{C_V^2 G_{\rm F}^2 \op^2}\,
\frac{Q_2}{Q_3}=
0.318\,\mu_{12}^2\,\left(\frac{10\,\rm keV}{\op}\right)^2\,\frac{Q_2}{Q_3}
\,,
\label{YYc}
\end{equation}
where $\mu_{12}\equiv\mu_\nu/10^{-12}(e/2m_e)$ and
\begin{equation}
\frac{Q_q}{Q_V}=\frac{q_\nu^2\,(4\pi\alpha)^2}{C_V^2 G_{\rm F}^2 \op^4}\,
\frac{Q_1}{Q_3}=
0.664\,q_{14}^2\,\left(\frac{10\,\rm keV}{\op}\right)^4\,\frac{Q_1}{Q_3}
\,,
\label{YYd}
\end{equation}
where $q_{14}\equiv q_\nu/10^{-14}$. In Fig.~7 we show contours of $Q_1/Q_3$
and $Q_2/Q_3$.

\section{Numerical Emission Rates}

Even though the methods described in Sect.~2 allow for a relatively simple
calculation of the plasma neutrino emission rates, one still needs to evaluate
several numerical integrals in order to determine the energy loss rate for
given values of density and temperature so that this procedure is not suitable
to be coupled directly with a stellar evolution code. However, we can use
these results to test the accuracy of widely used analytic approximation
formulae.

It turns out that for the plasma process the analytic approximations of
Beaudet, Petrosian, and Salpeter (1967), Munakata, Kohyama, and Itoh (1985),
and Schinder et~al.\ (1987) all agree with each other to an astonishing
accuracy if the same value for $C_V^2$ is used. In Fig.~8a we compare the
numerical rates of Beaudet et al. (1967) with the exact results obtained by
the Braaten and Segel method. We show contours for the relative deviation in
percent, $Q_{\rm analytic}^{\rm tot}/Q_{\rm exact}^{\rm tot}-1$.
We stress that what is plotted is the error of the {\it total\/} emission
rate; the plasma process alone may be more inaccurate than shown in regions
where it does not dominate. In this and the following figures we have used the
rates of Itoh et al.\ (1989) for the other emission processes. Thus,
$Q_{\rm analytic}^{\rm tot}$ and $Q_{\rm exact}^{\rm tot}$ differ only in the
treatment of the plasma process.  From Fig.~8a it is evident that the
numerical rates are rather poor approximations almost everywhere.
Nevertheless, it is these rates that have been used in virtually all stellar
evolution calculations.

Recently Itoh et al. (1992) have published numerical rates for the plasma
process which are claimed to be more accurate than 5\% wherever the plasma
process dominates. In Fig.~8b we put this claim to a test and find that there
remain substantial regions with much larger errors. We have checked that at
least some of these deviations also occur between their tabulated emission
rates and their fitting formula.

Blinnikov and Dunina-Barkovskaya (1992) have also derived a new analytic
approximation which is optimized in the region of small temperatures and
densities relevant for low-mass stars. We compare their rates with the exact
ones in Fig.~8c. Indeed, these rates are rather good fits for $T\lapprox 10^8$
and $\rho\lapprox10^6$, but for higher $T$ or $\rho$ they are unrelated to the
exact results. These approximation formulae involve multiplicative factors
which depend on $T$ alone and which approach 1 for $T\to0$. If we set these
factors to 1 for all $T$ the errors of the resulting simplified emission rates
are shown in Fig.~8d. The fit is not worse in the low-$T$ and low-$\rho$
region, and much better otherwise!

As we find none of the published fitting formulae satisfactory we have derived
a new one. To this end we have started with the $T=0$ version of Blinnikov and
Dunina-Barkovskaya (1992) and then ``flattened'' the errors with an extra
correction factor $f_{xy}$. As a result we have come up with
$Q_{\rm plas}=C_V^2\,Q_{\rm approx}$ (in $\rm erg\,cm^{-3}\,s^{-1}$) where
\begin{equation}
Q_{\rm approx}=3.00{\times}10^{21}\,\lambda^9\,\gamma^6\,e^{-\gamma}\,
(f_T+f_L)\,f_{xy}\,,
\end{equation}
where $\lambda=T/m_e$ and $\gamma=\omega_0/T$ where $\omega_0$ is the
zero-temperature plasma frequency, $\omega_0^2=4\pi\alpha N_e/E_{\rm F}$.
Numerically,
\begin{eqnarray}
\lambda&=&1.686{\times}10^{-10}\,T\nonumber\\
\gamma^2&=&\frac{1.1095{\times}10^{11}\,\rho/\mu_e}{T^2\,
[1+(1.019{\times}10^{-6}\,\rho/\mu_e)^{2/3}]^{1/2}}\,,
\end{eqnarray}
with $T$ in K, $\rho$ in $\rm g/cm^3$, and $\mu_e$ the number of baryons per
electron. Moreover, we have
\begin{eqnarray}
f_T&=&2.4+0.6\,\gamma^{1/2}+0.51\,\gamma+1.25\,\gamma^{3/2}\,, \nonumber \\
f_L&=&\frac{8.6\,\gamma^2+1.35\,\gamma^{7/2}}{225-17\,\gamma+\gamma^2}\,.
\end{eqnarray}
The coefficients here are slightly different from those used by Blinnikov and
Dunina-Barkovskaya (1992). Finally, we define
\begin{eqnarray}
x&=&\frac{1}{6}\Bigl[+17.5+\log_{10}(2\rho/\mu_e)-3\log_{10}(T)\Bigr]\,,
\nonumber\\
y&=&\frac{1}{6}\Bigl[-24.5+\log_{10}(2\rho/\mu_e)+3\log_{10}(T)\Bigr]\,,
\end{eqnarray}
where $x$ is a coordinate transverse to the diagonal band in Fig.~1 where the
plasma process is important, and $y$ is along this band. If $\vert x\vert>0.7$
or $y<0$ we use $f_{xy}=1$ and otherwise
\begin{eqnarray}
f_{xy}&\!=\!&1.05+\Bigl[0.39-1.25\,x-0.35\,\sin(4.5\,x)-0.3\,
\exp\{-(4.5\,x+0.9)^2\}\Bigr]{\times}\nonumber\\
&&\hskip1.5cm\times\,\exp\left\{-\left[\frac{\min(0,y-1.6+1.25\,x)}
      {0.57-0.25\,x}\right]^2\right\}\,.
\end{eqnarray}
We show the errors of our fitting formula in Fig.~8e; it is better than 5\%
everywhere and better than 4\%\ almost everywhere.

\section{Discussion and Summary}

In low-mass stars helium ignites in the cores of red giant stars under
degenerate conditions. Helium burning depends extremely sensitively on
temperature and density so that even relatively minor changes in the neutrino
cooling rates of the core change the ignition point and thus the core mass
$\M_{\rm tip}$ at the tip of the red giant branch, which in turn affects the
luminosity during the subsequent horizontal branch evolution. All previous
calculations of the core mass at helium flash seem to have used the Beaudet
et al.\ (1967), the Munakata et al.\ (1985), or the Schinder et al.\ (1987)
rates, all of which are equivalent with regard to the plasma process and thus
underestimate the emission rate as shown in Fig.~8a.

In order to illustrate the relevant range of parameters we show in Fig.~9 the
evolution of the central density and temperature of a $0.80\,\M_\odot$ star
where the tail ends of the arrows mark the conditions when the surface
brightness is at the indicated magnitudes. In Fig.~10 we show the red giant
part of this track overlaid with the errors of the Munakata et al.\ (1985)
rates that were used in our previous calculations (Raffelt and Weiss 1992).
The error contours are virtually the same as those shown in Fig.~8a. Near the
helium flash the average neutrino emission rate was underestimated by around
15\%.

Raffelt and Weiss (1992) found that the core mass at the helium flash varies
approximately as $\delta\M_c=0.020\delta F_\nu$ if the standard neutrino loss
rates are multiplied with a factor $F_\nu$, a result which agrees with the
previous calculations of Sweigart and Gross (1978). Therefore, the core mass
at the helium flash will be increased by about $0.004\,\M_\odot$.
In order to confirm this estimate we have re-calculated run~11 of Raffelt and
Weiss (1992) with the analytic emission rates derived in this paper. For
$\M=0.80\M_\odot$, $Z=10^{-4}$, and $Y_0=0.22$ we previously found
$\M_{\rm tip}=0.497\M_\odot$ for the core mass at helium ignition (at the tip
of the red giant branch) while we now find $\M_{\rm tip}=0.503\M_\odot$.
Therefore, we find an increase of $\delta\M_{\rm tip}=0.006\M_\odot$, in
reasonable agreement with our simple estimate. Thus, using the correct
neutrino emission rates changes the core mass at the helium flash by a small
but noticable amount.

In order to constrain neutrino dipole moments Raffelt and Weiss (1992) as
well as Blinnikov and Dunina-Barkovskaya (1992) and Castellani and
Degl'Innocenti (1992) have included the increased plasma losses in
evolutionary calculations by scaling the standard rates with a certain
factor. Instead of the exact ratio given by Eq.~(\ref{YYc}) they used
$Q_2/Q_3=1$ and $\op^2=\omega_0^2=4\pi\alpha N_e/E_{\rm F}$ which is the
zero-temperature value for the plasma frequency.  In Fig.~11 we show the
evolutionary track of Fig.~9 in the plane of $\gamma=\op/T$ and $v_*$.
Evidently on the RGB the interior of the core has an almost constant value
$\gamma\approx3$. A comparison with the upper panel of Fig.~7 reveals that by
using $Q_2/Q_3=1$ Raffelt and Weiss (1992) have underestimated the
dipole-induced emission rate by about 5\%.  An additional small error occurs
by using $\omega_0$ instead of $\op$. In Fig.~12 (upper panel) we show the
error of the dipole-induced emission rate if it had been scaled with the
correct standard rate.  These errors are so small that this scaling procedure
remains well justified.

In Fig.~12 (lower panel) we show the compound error of the dipole-induced
emission rate when coupled with the Munakata et al. (1985) emission rates.
Near helium ignition the emission rate was underestimated by $15-20$\%.
Because the core-mass and brightness at the helium flash vary approximately
linearly with $\mu_\nu$ in the range of interest, for a given value of
$\mu_\nu$ these quantities are changed by about 10\% more than given in
Raffelt and Weiss (1992). This is a negligible change with regard to bounds
on $\mu_\nu$.

The bounds on neutrino dipole moments discussed in Raffelt (1990) and Raffelt
and Weiss (1992) crudely amount to the constraint that the total neutrino
luminosity must not exceed about twice its standard value in order to
maintain the beautiful agreement between the observed and calculated
properties of globular cluster stars. We have already discussed that on the
RGB we may set $Q_2/Q_3=1$ in Eq.~(\ref{YYc}) without introducing a large
error, and similarly we may set $Q_1/Q_3=1$ in Eq.~(\ref{YYd}), although the
error here is somewhat larger. In any case, these approximations lead to an
underestimation of the non-standard emission rates and thus to conservative
bounds. Moreover, near the helium flash the density is about
$10^6\rm\, g/cm^3$ (see Fig.~10) so that $\op^2=17.8\,\rm keV$ in the center
of the star.  Therefore, in the center of the star we have for the total
energy-loss rate from Eqs.~(\ref{YYc}) and~(\ref{YYd})
\begin{equation}
Q_{\rm tot}/Q_V=1+1.07\,C_{V,R}^2+0.100\,\mu_{12}^2+0.066\,q_{14}^2\,.
\end{equation}
Away from the center the density and thus $\op$ is lower so that the
coefficients of $\mu_{12}^2$ and of $q_{14}^2$ would be larger if averaged
properly over the core. Hence this expression, again, is a conservative
estimate of the non-standard neutrino losses. With
$Q_{\rm tot}/Q_V<2$ as a formal criterion we find the bounds
\begin{eqnarray}
C_{V,R}&<&0.9\,,\nonumber\\
\mu_\nu&<&3{\times}10^{-12}\,(e/2m_e)\,,\nonumber\\
q_\nu&<&4{\times}10^{-14}\,.
\end{eqnarray}
In the model of Fukugita and Yanagida (1990) one has both a dipole moment and
r.h.\ interactions for the $\nu_e$'s so that in that model the constraints on
the dipole moment are more restrictive. The bound on $\mu_\nu$ alone is
virtually the same as that from the more detailed analysis of Raffelt (1990)
and Raffelt and Weiss (1992). The bound on $q_\nu$ is slightly more
restrictive than that found by Bernstein, Ruderman, and Feinberg (1963).

In summary, we have discussed in detail the neutrino emission rates from the
plasma process due to standard and non-standard interactions. By means of
Braaten and Segel's (1993) representation of the QED dispersion relations we
have tested the accuracy of widely used analytic approximation formulae, none
of which are found satisfactory. For the first time we have derived an
approximation formula which is accurate on the 4\% level wherever the plasma
process dominates. The correct emission rate leads to a slightly increased
core mass at the helium flash in low-mass stars, and to a slightly increased
sensitivity to non-standard neutrino losses. While these changes are
noticable they are so small that previous bounds on neutrino dipole moments
remain virtually unchanged.

\newpage

\section*{References}

{\parindent=0pt \frenchspacing

Adams, J.~B., Ruderman, M.~A., \& Woo, C.-H. 1963,
    Phys. Rev., 129, 1383.\bigskip

Beaudet, G., Petrosian, V., \& Salpeter, E.~E. 1967,
    Ap.~J., 150, 979.\bigskip

Bernstein, J., Ruderman, M., \& Feinberg, G. 1963, Phys. Rev., 132, 1227.
    \bigskip

Blinnikov, S.~I., \&\ Dunina-Barkovskaya, N.~V. 1992,
    Report MPA 861, submitted to M.N.R.A.S. \bigskip

Braaten, E. 1991, Phys. Rev. Lett., 66, 1655.\bigskip

Braaten, E., \&\ Segel, D. 1993, Report NUHEP-TH-93-1. \bigskip

Castellani, V., \& Degl'Innocenti, S. 1992, Report, Univ. of Pisa, to be
published.\bigskip

Dicus, D.~A. 1972, Phys. Rev. D, 6, 961.\bigskip

Fukugita, M., \&\ Yanagida, T. 1990,
    Phys. Rev. Lett., 65, 1975.\bigskip

Itoh, N., Adachi, T., Nakagawa, M., Kohyama, Y., \& Munakata, H. 1989,
    Ap.~J., 339, 354. (E) 1990, Ap.~J., 360, 741.\bigskip

Itoh, N., Mutoh, H., Hikita, A., \& Kohyama, Y. 1992,
    Ap.~J., 395, 622. \bigskip

Munakata, H., Kohyama, Y., \& Itoh, N. 1985,  Ap.~J., 296, 197.
    (E) 1986, Ap.~J., 304, 580.\bigskip

Nieves, J., \& Pal, P.~B. 1989a, Phys. Rev.~D, 39, 652.\bigskip

Nieves, J., \& Pal, P.~B. 1989b, Phys. Rev.~D, 40, 1350.\bigskip

Raffelt, G. 1990, Ap.~J., 365, 559.\bigskip

Raffelt, G., \&\ Weiss, A. 1992,
    Astron. Astrophys., 264, 536. \bigskip

Schinder, P.~J., Schramm, D.~N., Wiita, P.~J., Margolis, S.~H., \& Tubbs,
    D.~L. 1987, Ap.~J., 313, 531.\bigskip

Sweigart, A.~V., \& Gross, P.~G. 1978, Ap.~J. Suppl., 36, 405.\bigskip

Weldon, H.~A. 1982, Phys. Rev. D, 26, 1394.\bigskip

Zaidi, M.~H. 1965, Nuovo Cim., 40A, 502.

}


\newpage

\section*{Figure Captions}

\subsection*{Figure~1}
Regions of density and temperature where the different neutrino emission
processes contribute more than 90\% of the total. $\mu_e$ is the mean number
of baryons per electron. The bremsstrahlung contribution is for helium. The
dashed lines are contours for the indicated values of $Q_L/Q_T$, i.e.\ the
contribution of the longitudinal relative to the transverse plasma process.

\subsection*{Figure~2}
Function $G(x)$ as defined in Eq.~(\ref{Xfff}).

\subsection*{Figure~3}
Contours of $\gamma=\op/T$ and $v_*$ as defined in Eq.~(\ref{Xff}).

\subsection*{Figure~4}
Contours for $\pi_{T,k}=\omega_{T,k}^2-k^2$ in units of $\op^2$ and for
$Z_{T,k}$.

\subsection*{Figure~5}
Contours for $\pi_{L,k}=\omega_{L,k}^2-k^2$ in units of $\op^2$ and for
$Z_{L,k}^*=Z_{L,k}\pi_{L,k}/\omega_{L,k}^2$. The $\pi_{L,k}=0$ contour
corresponds to $\kmax(v_*)$ of Eq.~(\ref{Xg}). In the cross-hatched region
longitudinal plasmons have a space-like four momentum and thus cannot decay.

\subsection*{Figure~6}
Contours for $Q^L_3/Q^T_3$ as defined in Eq.~(\ref{YYb}). This ratio is
identical to the ratio of the longitudinal and transverse emission rates in
the standard model.

\subsection*{Figure~7}
Contours for $Q_2/Q_3$ and $Q_1/Q_3$ as defined in Eq.~(\ref{YYb}).

\subsection*{Figure~8}
(a)--(e) Comparison of the analytic plasma emission rates of the indicated
authors with the exact rates. Except for (e) the contours are at multiples of
$\pm10$\%. The errors in (a) of the Beaudet et al.\ (1967) rates are the same
for Munakata et al.\ (1985) and Schinder et al.\ (1987).

\subsection*{Figure~9}
Evolution of the central density and temperature of a $0.80\,\M_\odot$ star up
the red giant branch (RGB), and then from the horizontal branch (HB) up the
asymptotic giant branch (AGB). The rear-ends of the arrows are at the
indicated values for the surface brightness.

\subsection*{Figure~10}
Error in \% of the standard neutrino emission rates used in the evolutionary
calculations of Raffelt and Weiss (1992). The evolutionary track is that of
Fig.~9.

\subsection*{Figure~11}
Evolutionary track of Fig.~9 in the $\gamma$-$v_*$-plane.

\subsection*{Figure~12}
Error in \% of the dipole moment emission rates used in the calculations of
Raffelt and Weiss (1992).
{\bf Upper panel:} Error relative to the standard rates.
{\bf Lower panel:} Compound error if combined with the analytic standard rates
of Munakata et al.\ (1985).

\end{document}